\documentclass[jetpl]{article}

\usepackage{enumerate}
\usepackage{amsfonts}
\usepackage{amsmath}
\usepackage{comment}
\usepackage{textcomp}
\usepackage{amsthm, tikz}
\usepackage{amssymb}
\usepackage{color}
\usepackage{graphicx}
\usepackage[matrix,arrow,curve]{xy}
\usepackage{array}
\usepackage{mathtools}

\def\be{\begin{eqnarray}}
\def\ee{\end{eqnarray}}
\def\nn{\nonumber}

\usepackage{tikz}
\usepackage{braids}

\def\Tr{{\rm Tr}\,}

\def\l[{\phantom.[}

\hoffset= - 1.0in         

\begin{document}

\title{\vspace{.1cm}{\Large {\bf  Quantum Racah matrices and
3-strand braids in irreps $R$ with $|R|=4$}\vspace{.2cm}}
\author{
{\bf A.Mironov$^{a,b,c,d}$}\footnote{mironov@lpi.ru; mironov@itep.ru},
\ {\bf A.Morozov$^{b,c,d}$}\thanks{morozov@itep.ru},
\ {\bf An.Morozov$^{c,d,e}$}\footnote{andrey.morozov@itep.ru},
\ \ and
 \ {\bf A.Sleptsov$^{b,c,d,e}$}\thanks{sleptsov@itep.ru}}
\date{ }
}

\maketitle

\vspace{-5.5cm}

\begin{center}
\hfill FIAN/TD-11/16\\
\hfill IITP/TH-07/16\\
\hfill ITEP/TH-09/16
\end{center}

\vspace{3.3cm}

\begin{center}
$^a$ {\small {\it Lebedev Physics Institute, Moscow 119991, Russia}}\\
$^b$ {\small {\it ITEP, Moscow 117218, Russia}}\\
$^c$ {\small {\it Institute for Information Transmission Problems, Moscow 127994, Russia}}\\
$^d$ {\small {\it National Research Nuclear University MEPhI, Moscow 115409, Russia }}\\
$^e$ {\small {\it Laboratory of Quantum Topology, Chelyabinsk State University, Chelyabinsk 454001, Russia }}

\end{center}

\vspace{.5cm}

\begin{abstract}
We describe the inclusive Racah matrices for the first non-(anti)symmetric
rectangular representation $R=[2,2]$ for quantum groups $U_q(sl_N)$.
Most of them have sizes $2$, $3$, and $4$ and are fully described by the
eigenvalue hypothesis.
Of two 6x6 matrices, one is also described in this
way, but the other one corresponds to the case of degenerate eigenvalues
and is evaluated by the highest weight method.
Together with the much harder calculation for $R=[3,1]$ in \cite{mmms31}
and with the new method to extract exclusive matrices ${\cal S}$ and $\bar {\cal S}$
from the inclusive ones, this completes the story of Racah matrices
for $|R|\leq 4$ and allows one to calculate and investigate the corresponding
colored HOMFLY polynomials for arbitrary 3-strand and arborescent knots.
\end{abstract}

\vspace{.5cm}

\section{Introduction}

The Racah matrices or $6j$-symbols play an important role in theoretical physics
since its very early days.
It is enough to say that a whole chapter is devoted to this subject in \cite{LL3}.
This reflects the significance of symmetries for description of nature,
but also provides an example of how difficult is the study of symmetries themselves.
In fact, evaluation of the Racah matrices is an old classical problem,
which is still far from its final solution.
The modern approach is to attack this kind of problems by considering
a physical (quantum field theory) model, which contains nothing else,
but the quantities of interest: in the present case a model, where the correlators
are made from the Racah matrices.
Such a theory is nowadays well known: it is either a $2d$ Wess-Zumino-Novikov-Witten (WZNW)
conformal theory \cite{WZNW} or a $3d$ Chern-Simons theory \cite{CS}, the observables in the latter case being known
as {\it knot polynomials} \cite{knotpols}.
Speaking in modern terms, the two theories are related by a kind of holographic duality,
which is, hence, a far-going generalization of this old archetypical example.

Difficulties with the Racah calculus are, therefore, the difficulties with evaluation
of knot polynomials or, what is nearly the same, of monodromies of the conformal blocks.
Both subjects are now under intensive study, and the purpose of this letter is to report
some recent progress.
Namely, using the newly developed highest weight method (HWM) of \cite{MMMS21},
we succeeded in evaluation of the {\it inclusive} (i.e. for all $Q\in R^{\otimes 3}$)
quantum Racah matrices ${\cal U}_Q$
\be
 \Big\{(R\otimes R)\otimes R \longrightarrow Q\Big\} \ \ \
\stackrel{{\cal U}_Q}{\longrightarrow} \ \ \
 \Big\{R\otimes (R\otimes R) \longrightarrow Q\Big\}
 \label{racah}
\ee
for all representations
$R$ of sizes up to $4$: the really new part concerns $R=[3,1]$ and $R=[2,2]$.
For symmetric representations $R=[1],[2],[3],[4]$, the answers are known from \cite{MMMkn12,IMMMev},
and are actually described by the eigenvalue hypothesis of \cite{IMMMev},
for antisymmetric $R=[1,1],[1,1,1],[1,1,1,1]$ they are obtained by
a change of variable $q\longrightarrow q^{-1}$, for $R=[21]$ they were found  in \cite{MMMS21}.
These ${\cal U}_Q$ are actually matrices, acting in the space of representations $Y\in R^{\otimes 2}$,
which appear in the first products $(R\otimes R)$ in (\ref{racah}) and contribute to a given $Q$.
For non-rectangular representations (i.e. when the Young diagram $R$ is not rectangular),
the representations $Y$ and $Q$ come with non-trivial multiplicities, which leads to certain
degeneracies and makes the problem really difficult.
A part of the problem is that in such cases the Racah {\it matrices} depend on the basis choices
and lack invariant (basis independent) definition, thus distracting pure mathematicians from such studies,
despite a severe need for explicit answers for purposes of quantum field and string theory.
We described this problem in some detail in recent \cite{mmms31}
and do not repeat the relevant details of the HWM here.
Instead in this short presentation, we concentrate on the case of $R=[2,2]$, which is rectangular
and free of these additional complications.
Our purpose is to enumerate ingredients of the calculations and immediate checks
performed after getting the answer.

The Racah matrices {\it per se} appear in expressions for the knot polynomials in two cases:
the exclusive matrices $(R,R,\bar R\longrightarrow R)$ describe arborescent (double fat) knots \cite{mmmrv,mmmsrv},
while the inclusive ones $(R,R,R\longrightarrow Q)$ the three-strand knots.
Exclusive are actually deducible from inclusive \cite{mmmsEI}, and here we concentrate on the latter ones.
Knots with more strands require complicated contractions of the Racah matrices
(called {\it mixing matrices} in \cite{MMMkn12}), which also turn to be much simpler and explorable than
they seem, but this is also beyond the scope of the present letter.

\section{Pattern of Racah matrices arising for the size $|R|=4$}

\subsection{Specification to the case of $R=[4]$}

This case of symmetric representation was exhaustively considered in \cite{MMMkn12,IMMMev} and can serve as a sample
for other representations. However, it is the simplest one, since all Racah matrices are given by the $U_q(sl_2)$ formulas.
Square of the symmetric representation $[4]$ is decomposed in a very simple way:
\be
[4]\otimes [4] = [8]_++[7,1]_- + [6,2]_+ + [5,3]_- + [4,4]_+
\ee
where subscript is plus or minus for the representations from the symmetric and antisymmetric
squares respectively.
The representation content of the cube is now
\be
[4]\otimes[4]\otimes[4] =
\,[12]+\,[6, 6]+3\,[7, 5]+5\,[8, 4]+4\,[9, 3]+3\,[10, 2]+2\,[11, 1]+\,[4, 4, 4]+2\,[5, 4, 3]+
\,[5, 5, 2]+ \nn \\
\!\!\!\!\!\!\!\!\!\!\!
+\,[6, 3, 3]+3\,[6, 4, 2]+2\,[6, 5, 1]+2\,[7, 3, 2]+4\,[7, 4, 1]+\,[8, 2, 2]+3\,[8, 3, 1]
+2\,[9, 2, 1]+\,[10, 1, 1]
\label{decocube}
\nn
\ee
and most items come with non-unit multiplicities.
However, these multiplicities are in one-to-one correspondence with the
intermediate representations $Y\in R^{\otimes 2}$.
From this, one can read off content of the inclusive Racah matrices:
{\footnotesize
\be
\begin{array}{|c|c|c|}
\hline
&&\text{number of}\\
\text{matrix size} & Q &  \\
&&\text{matrices}\\ \hline && \\
1 & [12], [10, 1, 1], [8, 2, 2], [6, 6], [6, 3, 3], [5, 5, 2], [4, 4, 4] & 7 \\
&&\\ \hline && \\
2 & [11, 1], [5, 4, 3], [6, 5, 1], [7, 3, 2],  [9, 2, 1] & 5 \\
&&\\ \hline && \\
3 & [10, 2], [8, 3, 1], [7, 5], [6, 4, 2] & 4 \\
&&\\ \hline && \\
4 & [9, 3], [7, 4, 1] & 2 \\
&&\\ \hline && \\
5 & [8, 4] & 1 \\
&&\\ \hline
 \end{array}
 \nn
\ee
}
The biggest matrix has size $5\times 5$.
This means that all Racah matrices arising in $4^{\otimes3}$ can be found with the help
of the eigenvalue hypothesis \cite{IMMMev}.

\subsection{Specification to the case of $R=[3,1]$
\label{hv31}}

This is a hard case, described in detail in \cite{mmms31}, where we refer the reader to.
The results are presented there in the same format, as in the present paper.

\subsection{Specification to the case of $R=[2,2]$}

This case is again relatively simple, still this is a new piece of knowledge.

Decomposition of the square
\be
\l[22]\otimes [22] = [44]\oplus [431]_-\oplus[422]\oplus [3311]\oplus [3221]_-\oplus [2222]
\ee
contains six irreducible representations.
Note that the symmetric diagrams $Y=[332]$ and $Y=[4211]$ do not contribute.

This implies that maximal size of the Racah mixing matrices will be $6\times 6$.
Another simplifying fact is that all representations come with no multiplicities,
moreover, all ${\cal R}$-matrix are different (there are no accidental coincidence,
which are often encountered for non-rectangular representations $R$).
Together these two facts give a chance that the mixing matrices can be defined
from eigenvalue hypothesis \cite{IMMMev}, which provides explicit expressions
for the entries of Racah matrices through eigenvalues of the ${\cal R}$-matrices
(they are given in \cite{IMMMev} for sizes up to $5$ and size $6$ was later described
in \cite{MMuniv}).

Representation content of the cube is
{\footnotesize
\be
[2,2]\otimes\Big([2,2]\otimes[2,2]\Big) &=&
3\,[4, 2, 2, 2, 2]+\,[4, 2, 2, 2, 1, 1]+\,[3, 3, 2, 2, 2]+3\,[3, 3, 2, 2, 1, 1]+2\,[3, 2, 2, 2, 2, 1]+\,[2, 2, 2, 2, 2, 2]+\nn \\
&+&4\,[5, 3, 2, 1, 1]+3\,[5, 3, 2, 2]+3\,[5, 3, 3, 1]+\,[3, 3, 3, 1, 1, 1]+2\,[6, 3, 2, 1]+3\,[4, 4, 3, 1]+\,[6, 3, 3]+\,[4, 4, 4]+\nn \\
&+&\,[6, 4, 1, 1]+2\,[5, 4, 1, 1, 1]+6\,[5, 4, 2, 1]+2\,[5, 4, 3]+3\,[5, 5, 1, 1]+\,[5, 5, 2]+2\,[3, 3, 3, 2, 1]+\,[3, 3, 3, 3]+\nn \\
&+&2\,[4, 3, 2, 1, 1, 1]+6\,[4, 3, 2, 2, 1]+3\,[4, 3, 3, 1, 1]+3\,[4, 3, 3, 2]+\,[6, 2, 2, 2]+\,[4, 4, 1, 1, 1, 1]+3\,[4, 4, 2, 1, 1]+\nn \\
&+&3\,[6, 4, 2]+2\,[5, 2, 2, 2, 1]+2\,[6, 5, 1]+\,[6, 6]+6\,[4, 4, 2, 2]
\nn
\ee
}
and the inclusive Racah matrices form the collection

{\footnotesize
\be
\begin{array}{|c|c|c|}
\hline
&&\text{number of}\\
\text{matrix size} & Q &  \\
&&\text{matrices}\\ \hline && \\
1 & [6{,} 6]{,} [6{,} 3{,} 3]{,} [6{,} 4{,} 1{,} 1]{,} [6{,} 2{,} 2{,} 2]{,} [5{,} 5{,} 2]{,} [4{,} 4{,} 4]{,} [4{,} 4{,} 1{,} 1{,} 1{,} 1]{,} [4{,} 2{,} 2{,} 2{,} 1{,} 1]{,} [3{,} 3{,} 3{,} 3]{,} [3{,} 3{,} 2{,} 2{,} 2]{,} [3{,} 3{,} 3{,} 1{,} 1{,} 1], [2{,} 2{,} 2{,} 2{,} 2{,} 2] & 12 \\
&&\\ \hline && \\
2 & [6{,} 5{,} 1]{,} [6{,} 3{,} 2{,} 1]{,} [5{,} 4{,} 3]{,} [5{,} 4{,} 1{,} 1{,} 1]{,} [5{,} 2{,} 2{,} 2{,} 1]{,} [4{,} 3{,} 2{,} 1{,} 1{,} 1]{,} [3{,} 3{,} 3{,} 2{,} 1]{,} [3{,} 2{,} 2{,} 2{,} 2{,} 1] & 8 \\
&&\\ \hline && \\
3 & [6{,} 4{,} 2]{,} [5{,} 5{,} 1{,} 1]{,} [5{,} 3{,} 3{,} 1]{,} [5{,} 3{,} 2{,} 2]{,} [4{,} 4{,} 3{,} 1]{,} [4{,} 3{,} 3{,} 2]{,} [4{,} 4{,} 2{,} 1{,} 1]{,} [4{,} 3{,} 3{,} 1{,} 1]{,} [4{,} 2{,} 2{,} 2{,} 2]{,} [3{,} 3{,} 2{,} 2{,} 1{,} 1] & 10 \\
&&\\ \hline && \\
4 & [5{,} 3{,} 2{,} 1{,} 1] & 1 \\
&&\\ \hline && \\
5 & - & 0 \\
&&\\ \hline && \\
6 & [5{,} 4{,} 2{,} 1]{,} [4{,} 4{,} 2{,} 2]{,} [4{,} 3{,} 2{,} 2{,} 1] & 3 \\
&&\\ \hline
 \end{array}
 \nn
\ee
}

\noindent
All matrices of size up to four are nicely handled by the eigenvalue hypothesis
(though we checked them by the direct highest weight calculation as well).
We do not list them here.
The $6\times 6$ matrix
{\footnotesize
\be
{\cal U}_{[4,4,2,2]} = \left(\begin{array}{cccccc}
\frac{[2]^2}{[5][4]^2} & \sqrt{\frac{[3]}{[5]}}\frac{[2]^2}{[4]^2} & \sqrt{\frac{1}{[5]}}\frac{[2]}{[4]} & -\sqrt{[7][3]}\frac{[3][2]}{[5][4]} & -\sqrt{[7]}\frac{[2]^2}{[4]^2} & \sqrt{\frac{[7][3]}{[5]}}\frac{[6][2]}{[4]^2[3]} \\ \\
-\sqrt{\frac{[3]}{[5]}}\frac{[2]^2}{[4]^2} & -\frac{[3][2]([7]+[3]-2)}{[6][4]^2} & \sqrt{[3]}\frac{[8][3]}{[6][4]^2} & \sqrt{\frac{[7]}{[5]}}\frac{[8][3]}{[6][4]^2} & \sqrt{[7][5][3]}\frac{[3][2]}{[6][4]^2} & \sqrt{[7]}\frac{[2]^2}{[4]^2} \\ \\
\sqrt{\frac{1}{[5]}}\frac{[2]}{[4]} & -\sqrt{[3]}\frac{[8][3]}{[6][4]^2} & \frac{[3]([9]-[5]+[3]+2)}{[6][5][2]} & \sqrt{\frac{[7][3]}{[5]}}\frac{[3]}{[6][2]} & \sqrt{\frac{[7]}{[5]}}\frac{[8][3]}{[6][4]^2} & \sqrt{[7][3]}\frac{[2]}{[5][4]} \\ \\
\sqrt{[7][3]}\frac{[2]}{[5][4]} & -\sqrt{\frac{[7]}{[5]}}\frac{[8][3]}{[6][4]^2} & \sqrt{\frac{[7][3]}{[5]}}\frac{[3]}{[6][2]} & \frac{[3]([9]-[5]+[3]+2)}{[6][5][2]} & \sqrt{[3]}\frac{[8][3]}{[6][4]^2} & -\sqrt{\frac{1}{[5]}}\frac{[2]}{[4]} \\ \\
\sqrt{[7]}\frac{[2]^2}{[4]^2} & \sqrt{[7][5][3]}\frac{[3][2]}{[6][4]^2} & -\sqrt{\frac{[7]}{[5]}}\frac{[8][3]}{[6][4]^2} & -\sqrt{[3]}\frac{[8][3]}{[6][4]^2} & -\frac{[3][2]([7]+[3]-2)}{[6][4]^2} & -\sqrt{\frac{[3]}{[5]}}\frac{[2]^2}{[4]^2} \\ \\
\sqrt{\frac{[7][3]}{[5]}}\frac{[6][2]}{[4]^2[3]} & -\sqrt{[7]}\frac{[2]^2}{[4]^2} & \sqrt{[7][3]}\frac{[3][2]}{[5][4]} & -\sqrt{\frac{1}{[5]}}\frac{[2]}{[4]} & \sqrt{\frac{[3]}{[5]}}\frac{[2]^2}{[4]^2} & \frac{[2]^2}{[5][4]^2}
\end{array}\right)
\ee}
was evaluated by the HWM, but {\it a posteriori} is described
by the eigenvalue formula \cite[eq.(17)]{MMuniv} with
\be
\lambda_{[4,4]} = q^{-8}, \  \lambda_{[4,3,1]} = -q^{-4}, \ \lambda_{[4,2,2]} = q^{-2}, \ \lambda_{[3,3,1,1]} = q^{2}, \ \lambda_{[3,2,2,1]} = -q^{4}, \ \lambda_{[2,2,2,2]} = q^{8}.
\ee
The remaining two $6\times 6$ matrices correspond to mutually transposed diagrams which are thus
related by the change $q\longrightarrow -q^{-1}$ (rank-level duality \cite{DMMSS,GS,IMMMfe}), so that only one of them needs to be
calculated.
In this case, there are two coincident eigenvalues, thus one expects it to have
a block form in an appropriate basis, but we did not yet manage to confirm this
natural expectation.
Instead we found it by brute force, with the HWM:

{\footnotesize
\be
{\cal U}_{[5,4,2,1]} = \left(\begin{array}{cccccc}
\frac{[2]}{[5][4]} & \sqrt{\frac{[6][2]}{[5]\alpha_0}}\frac{[3]}{[4]} & -\sqrt{\frac{[8][3]}{[5][4]\alpha_0}}\frac{[2]}{[4]} & -\sqrt{\frac{1}{[5]}} & -\sqrt{\frac{[8][3]}{[4]}}\frac{1}{[5]} & \sqrt{\frac{[8]}{[5][4]}} \\ \\
\sqrt{\frac{[6][2]}{[5]\alpha_0}}\frac{[3]}{[4]} & \frac{[3]}{[2]^2\alpha_0} & -\sqrt{\frac{[8][3]}{[6][4][2]}}\frac{\alpha_1}{[4]\alpha_0} & \sqrt{\frac{1}{[6][2]\alpha_0}}\frac{[4][3]}{[2]^2} & -\sqrt{\frac{[8][3][2]\alpha_0}{[6][5][4]}}\frac{1}{[2]^2} & -\sqrt{\frac{[8][2]}{[6][4]\alpha_0}}\frac{[5][3]}{[4][2]} \\ \\
-\sqrt{\frac{[8][3]}{[5][4]\alpha_0}}\frac{[2]}{[4]} & -\sqrt{\frac{[8][3]}{[6][4][2]}}\frac{\alpha_1}{[4]\alpha_0} & \frac{[3]\alpha_2}{[6][4]\alpha_0} & \sqrt{\frac{[8][2]}{[4]\alpha_0}}\frac{[3]}{[4][2]} & -\sqrt{\frac{\alpha_0}{[5]}}\frac{[3]}{[6][2]} & \sqrt{\frac{[3]}{\alpha_0}}\frac{[7][3]}{[6][4]} \\ \\
-\sqrt{\frac{1}{[5]}} & \sqrt{\frac{1}{[6][2]\alpha_0}}\frac{[4][3]}{[2]^2} & \sqrt{\frac{[8][2]}{[4]\alpha_0}}\frac{[3]}{[4][2]} & -\frac{[3]^2-3[3]}{[6][2]} & -\sqrt{\frac{[8][4][3]}{[5]}}\frac{[3]}{[6][2]^2} & -\sqrt{\frac{[8]}{[4]}}\frac{[3]}{[6][2]} \\ \\
-\sqrt{\frac{[8][3]}{[4]}}\frac{1}{[5]} & -\sqrt{\frac{[8][3][2]\alpha_0}{[6][5][4]}}\frac{1}{[2]^2} & -\sqrt{\frac{\alpha_0}{[5]}}\frac{[3]}{[6][2]} & -\sqrt{\frac{[8][4][3]}{[5]}}\frac{[3]}{[6][2]^2} & -\frac{[8][3]}{[6][5][4][2]} & -\sqrt{\frac{[3]}{[5]}}\frac{[3]}{[6][2]} \\ \\
\sqrt{\frac{[8]}{[5][4]}} & -\sqrt{\frac{[8][2]}{[6][4]\alpha_0}}\frac{[5][3]}{[4][2]} & \sqrt{\frac{[3]}{\alpha_0}}\frac{[7][3]}{[6][4]} & -\sqrt{\frac{[8]}{[4]}}\frac{[3]}{[6][2]} & -\sqrt{\frac{[3]}{[5]}}\frac{[3]}{[6][2]} & -\frac{[3]}{[6][4]}
\end{array}\right)
\ee}
with $\alpha_2=[11]-[7]+1, \ \alpha_1=[9]+2[7]+2[5]+[3]-1, \ \alpha_0=[7]+[5]-1$.

\bigskip

We are now ready to consider a particularly important application of these results
on non-trivial Racah matrices: to the {\it colored} knot polynomials.

\section{[22]-colored polynomials and their properties}

\subsection{The 2- and 3-strand HOMFLY polynomial}

The basic formula in the theory of knot polynomials, revealing
their group theory nature is the one for the
HOMFLY polynomial of the 2-strand braid with $m$ crossings (which describes the torus knot/link),
colored by the representation (Young diagram) $R$:
\be
H_R^{(m)} = \sum_{Y\in R\otimes R} \frac{d_Y}{d_R} \cdot
\left(\frac{\epsilon_Y q^{\varkappa_Y}}{q^{4\varkappa_R}A^{|R|}}\right)^m
\label{2straH}
\ee
with
$\epsilon_Y=\pm 1$ depending on whether $Y$ belongs to the symmetric or antisymmetric
square of $R$, and with $q$ to the power of Casimir eigenvalue $\varkappa_Y = \sum_{(i,j)\in Y} \big(i-j\big)$
representing the eigenvalues of quantum ${\cal R}$-matrix in the representation $Y$
of quantum dimension
\be
d_Y = {\rm dim}_q(Y)={\rm Schur}_Y\left(p_k=\{A^k\}/\{q^k\}\right)=
\prod_{(i,j)\in Y} \frac{\{Aq^{i-j}\}}{\{q^{1+{\rm arm}(i,j)+{\rm leg}(i,j)}\}},
\ee
As usual, $\{x\} = x-x^{-1}$.
For multi-strand braids the {\it evolution factor} in power $m$ is substituted by a more
complicated element of the braid group, expressed via the Racah matrices intertwining ${\cal R}$ that
act on different pairs of the adjacent braids.
According to \cite{MMMkn12}, the link/knot polynomial for a closure of the 3-strand braid
$B^{m_1,n_1|m_2,n_2|\ldots}$  is equal to
\be
H_R^{(m_1,n_1|m_2,n_2|\ldots)} = \sum_{Q\in R^{\otimes 3}}
\ \frac{d_Q}{d_R}\cdot
\Tr_Q  \Big\{ {\cal R}_Q^{m_1}{\cal U}_Q {\cal R}_Q^{n_1}{\cal U}^\dagger_Q
{\cal R}_Q^{m_2}{\cal U}_Q {\cal R}_Q^{n_2}{\cal U}^\dagger_Q \ldots\Big\}
\label{3strafla}
\ee
In the following picture $m_1=0,n_1= -2,m_2=2,n_2=-1,m_3=3$:

\bigskip

\unitlength 0.8mm 
\linethickness{1pt}
\ifx\plotpoint\undefined\newsavebox{\plotpoint}\fi 
\begin{picture}(145.5,53)(-30,0)
\put(19.5,34.5){\line(1,0){13.25}}
\put(41.25,43.25){\line(1,0){11.25}}
\put(19.25,43){\line(1,0){13.25}}
\put(38.75,35){\line(1,0){13.75}}
\put(61.25,43.25){\line(1,1){8.75}}
\put(70,52){\line(1,0){14.75}}
\put(18.5,52){\line(1,0){41}}
\multiput(59.5,52)(.033505155,-.043814433){97}{\line(0,-1){.043814433}}
\put(58.25,35.25){\line(1,0){33.75}}
\multiput(92,35.25)(.033505155,.038659794){97}{\line(0,1){.038659794}}
\multiput(64.5,45)(.03289474,-.04605263){38}{\line(0,-1){.04605263}}
\put(65.75,43.25){\line(1,0){19}}
\multiput(84.5,43.5)(.0346153846,.0336538462){260}{\line(1,0){.0346153846}}
\multiput(84.75,52)(.03370787,-.03651685){89}{\line(0,-1){.03651685}}
\multiput(52.5,43)(.033653846,-.046474359){156}{\line(0,-1){.046474359}}
\multiput(52.5,35)(.03353659,.03353659){82}{\line(0,1){.03353659}}
\multiput(56.75,39)(.035447761,.03358209){134}{\line(1,0){.035447761}}
\multiput(32.25,43)(.033602151,-.041666667){186}{\line(0,-1){.041666667}}
\multiput(32.75,34.75)(.03333333,.03333333){75}{\line(0,1){.03333333}}
\put(37,39){\line(1,1){4.25}}
\put(99.75,35.25){\line(1,0){45.75}}
\multiput(100,35.5)(-.0336990596,.0352664577){319}{\line(0,1){.0352664577}}
\multiput(97.25,41)(.0336363636,.04){275}{\line(0,1){.04}}
\put(106.5,52){\line(1,0){7.75}}
\put(121.25,44){\line(1,0){6.75}}
\put(128,44){\line(5,6){7.5}}
\put(135.5,53){\line(1,0){8.25}}
\put(93.25,52.25){\line(1,0){5.75}}
\multiput(99,52.25)(.03353659,-.04268293){82}{\line(0,-1){.04268293}}
\multiput(103,47)(.03333333,-.05){60}{\line(0,-1){.05}}
\put(105,44){\line(0,1){0}}
\put(105,44){\line(1,0){9.5}}
\multiput(114.5,44)(.033632287,.036995516){223}{\line(0,1){.036995516}}
\put(122,52.25){\line(1,0){5.25}}
\multiput(127.25,52.25)(.03353659,-.03963415){82}{\line(0,-1){.03963415}}
\multiput(131.5,47)(.03333333,-.04166667){60}{\line(0,-1){.04166667}}
\put(133.5,44.5){\line(1,0){10.75}}
\multiput(114.25,52.25)(.03370787,-.03651685){89}{\line(0,-1){.03651685}}
\multiput(121,44)(-.03333333,.04666667){75}{\line(0,1){.04666667}}
\label{3strand}
\end{picture}

\vspace{-2.4cm}

Actually, (\ref{2straH}) is a particular case of (\ref{3strafla})
for the one-parametric subfamily $(m,0)$, and evolution in any parameter
$n$ can be described by its simple generalization:
\be
H_R^{(m,n_1|m_2,n_2|\ldots)} = \sum_{Y\in R\otimes R} \frac{d_Y}{d_R} \cdot
\left(\frac{\epsilon_Y q^{\varkappa_Y}}{q^{4\varkappa_R}A^{|R|}}\right)^m \cdot
C_{R,Y}^{(\cdot,n_1|m_2,n_2|\ldots)}(A,q)
\label{1famevo}
\ee
with correction coefficients $C_{R,Y}$ depending on all other parameters of the braid.
Despite they are not just unity as in (\ref{2straH}), they often have rather simple
and comprehensible structure, see \cite{mmms31} and eq.(\ref{C}) below for examples.
A similar formula for the double-parameter evolution
\be
H_R^{(m_1,n_1|m_2,n_2|\ldots)} = \sum_{Y_1,Y_2\in R\otimes R}
\left(\frac{\epsilon_{Y_1} q^{\varkappa_{Y_1}}}{q^{4\varkappa_R}A^{|R|}}\right)^{m_1}
\left(\frac{\epsilon_{Y_2} q^{\varkappa_{Y_2}}}{q^{4\varkappa_R}A^{|R|}}\right)^{m_2}\cdot
h_{R,Y_1,Y_2}^{(\cdot,n_1|\cdot,n_2|\ldots)}(A,q)
\label{2famevo}
\ee
plays the central role in the ${\cal U} - {\cal S}$ relation of \cite{mmmsEI}.

\subsection{Specification to $R=[2,2]$}

The newly calculated Racah matrices allow us to consider the case $R=[2,2]$.
So far the HOMFLY polynomials were available in this representation only for peculiar torus
knots, where the Rosso-Jones formula \cite{RJ,DMMSS} provides an exhaustive
generalization of (\ref{2straH}).
For a new list of $H_{[22]}$ for all knots from the Rolfsen table (i.e. up to 10 crossings) that have 3-strand braid representation
see \cite{knotebook}.
They all satisfy the factorization properties \cite{DMMSS,IMMMfe,chi}
\be
H_R(q=1,A) \ =\  H_{[1]}(q=1,A)^{|R|}
\ee
and \cite{Konfact}
\be
H_{[22]}=H_{[31]}=H_{[4]} = H_{[2]}^2  \ \ \ \ &{\rm at} \ \ q^4=1 \nn \\
H_{[22]}=H_{[4]} \ \ \ \ &{\rm at} \ \ q^6=1
\ee
Note that, since $R=[2,2]$ is not a single-hook diagram,
there is {\it no} simple statement about the colored Alexander polynomial at $A=1$.

Also, the properties of the perturbative \cite{Kont} and genus (Hurwitz) \cite{MMS} expansions,
described in sec.6 of \cite{mmms31} remain true,
though the checks are limited by the same reasons as in that paper.

As to the differential expansion \cite{DGR,IMMMfe,evo,arthdiff}, it is an interesting separate issue.
For the rectangular diagrams $R=[r^s]$, the elementary representation theory predicts that
\be
H^{\cal K}_{[r^s]} -1 \sim \{Aq^r\}\{A/q^s\}
\ee
but the next terms are not so straightforward.
We give here just a simple example for the figure eight knot,
which was the first origin of insights in this topic for symmetric representations \cite{IMMMfe}:
\be
\!\!\!\!\!\!
H_{[22]}^{4_1} = 1 + \{Aq^2\}\{A/q^2\}\Big\{[2]^2 - [3]\Big(\{Aq^3\}\{A/q\}+\{Aq\}\{A/q^3\}\Big)
+ \nn \\
+ [2]^2\{Aq^3\}\{Aq\}\{A/q\}\{A/q^3\} + \{Aq^3\}\{Aq^2\}\{Aq\}\{A/q\}\{A/q^2\}\{A/q^3\}\Big\}
\ee

The last thing we do in this letter
is a concise description of $H_{[22]}$ for a simple, but important sub-family of 3-strand braids.

\subsection{Evolution for $(m,-1|1,-1)$}

This family contains a number of interesting prime knots:
$4_1,5_2,6_2,7_3,8_2,9_3,10_2,\ldots$, corresponds to the pretzel knots $(m,\bar 2,1)$ \cite{MMSpret,mmms31}
and is described by the formula (\ref{1famevo}).
Our knowledge of the Racah matrices allows us to evaluate the correction coefficients $C$
for $R=[2,2]$ (we keep only diagram index $Y$ to simplify the formulas):
\be
H_{[2,2]}^{(m,-1\,|\,1,-1)}
= \frac{A^{-4m}}{\{q\}^4d_{[22]}}\!\!\!\!\!\!\!\!\!\!\!\!\!\!\!\!\!\!
&\Big(d_{[44]}C_{[44]}\cdot q^{8m} + d_{[431]}C_{[431]}\cdot (-q^4)^m +
d_{[422]}C_{[422]}\cdot q^{2m} + \nn \\
&+  d_{[3311]}C_{[3311]}\cdot q^{-2m} + d_{[3221]}C_{[3221]}\cdot (-q^{-4})^m
+ d_{[2222]}C_{[2222]}\cdot q^{-8m}\Big)
\ee
and
\be\label{C}
\begin{array}{cccccc}
C_{[44]} =& A^8 &- q^2[2]^2A^6\{q\} &+ q^4[3][2]A^4\{q\}^2\cdot (q+q^{-1}[3])
&-q^{6}[3][2]^3A^2\{q\}^3&+q^8[3][2]^2\{q\}^4\nn \\
C_{[431]} =& A^8 &- q^2[2]A^6\{q\}([3]-q^{-4})& + q^2[3]^2[2]A^4\{q\}^2\cdot (q-q^{-1} )
&-q^{2}[3][2]A^2\{q\}^3(q^4-[3])&-q^4[3][2]^2\{q\}^4\nn \\
C_{[422]} = &A^8 &+ q^{-2}[2]A^6\{q\}(1-q^5[2]) &+ q^4[3][2]A^4\{q\}^2\cdot (q-q^{-2}[2] )
&+q^{4}[3][2]A^2\{q\}^3(1-q^{-5}[2])&+q^2[3][2]^2\{q\}^4\nn \\
C_{[3311]} =& A^8 &- q^2[2]A^6\{q\}(1-q^{-5}[2]) &+ q^{-4}[3][2]A^4\{q\}^2\cdot (q^{-1}-q^{2}[2] )
&-q^{-4}[3][2]A^2\{q\}^3(1-q^{5}[2])&+q^{-2}[3][2]^2\{q\}^4\nn \\
C_{[3221]} =& A^8 &+ q^{-2}[2]A^6\{q\}([3]-q^{4})& + q^2[3]^2[2]A^4\{q\}^2\cdot (q^{-1}-q )
&+q^{-2}[3][2]A^2\{q\}^3(q^{-4}-[3])&-q^{-4}[3][2]^2\{q\}^4\nn \\
C_{[222]} = &A^8 &+ q^{-2}[2]^2A^6\{q\}& + q^{-4}[3][2]A^4\{q\}^2\cdot (q^{-1}+q [3])
&+q^{-6}[3][2]^3A^2\{q\}^3&+q^{-8}[3][2]^2\{q\}^4
\end{array}
\ee
For a similar result for the (anti)symmetric representations and for $R=[3,1]$ see
\cite{mmms31}, for extension from (\ref{1famevo}) to   (\ref{2famevo}) see \cite{mmmsEI}.

\section{Conclusion}

To conclude, we reported the results of inclusive Racah matrix calculation for
representation $R=[2,2]$.
This is a nice example, where calculations are already complicated, but the outcome
is relatively concise and can be presented in a short communication.
Not only being a step forward in solving a long standing problem in theoretical physics,
the result has an immediate application to the study of colored knot polynomials
which generalize conformal blocks and are in the center of attention in modern studies.
The explicit calculation confirms some of the existing conjectures in this field and
poses new questions about the genus, Hurwitz and differential expansions,
important subjects in many branches of quantum field theory in various dimensions.

\section*{Acknowledgements}

This work was funded by the Russian Science Foundation (Grant No.16-11-10291).

\end{document}